\documentclass[prl,twocolumn]{revtex4}
\usepackage{graphicx}
\usepackage{dcolumn}
\usepackage{multirow}
\usepackage{amsmath,amsthm,amssymb}
\usepackage{subfigure}

\def\ra{\rightarrow}
\def\ep{\varepsilon}
\newcommand{\inner}[2]{\left( #1 \left| #2 \right. \right)}
\newcommand{\coef}[1]{\left[\ep^{#1}\right]}

\begin{document}
\title{Time-periodic solutions in Einstein AdS - massless scalar field system}

\author{Maciej Maliborski}
\email{maliborski@th.if.uj.edu.pl}
\affiliation{M. Smoluchowski Institute of Physics, Jagiellonian University, Krak\'ow, Poland}

\author{Andrzej Rostworowski}
\email{arostwor@th.if.uj.edu.pl}
\affiliation{M. Smoluchowski Institute of Physics, Jagiellonian University, Krak\'ow, Poland}
\date{\today}
\begin{abstract}
We construct time-periodic solutions for a system of self-gravitating massless scalar field, with negative cosmological constant, in $d+1$ spacetime dimensions at spherical symmetry, both perturbatively and numerically. We estimate the convergence radius of the formally obtained perturbative series and argue that it is greater then zero. Moreover, this estimate coincides with the boundary of the convergence domain of our numerical method and the threshold for the black-hole formation. Then we confirm our results with a direct numerical evolution. This also gives strong evidence for nonlinear stability of the constructed time-periodic solutions.
\end{abstract}

\maketitle

\textit{Introduction}. A recent numerical and analytical study of the four dimensional spherically symmetric Einstein-massless scalar field equations with negative cosmological constant indicated that anti-de Sitter (AdS) space is unstable against the formation of a black hole under arbitrarily small generic perturbations \cite{br}. Qualitatively the same results were  obtained later in higher dimensions \cite{jrb, bll}. Although gravitational collapse seems to be a generic fate of a small perturbation of AdS, it was suggested in \cite{br} that there may exist non-generic initial data for which the evolution remains globally regular in time.  This conjecture was later generalized to the vacuum Einstein equations with negative cosmological constant \cite{dhs, dhms}, where time-periodic solutions (\textit{geons}) were postulated. In this note, continuing the study initiated in \cite{br}, we give strong evidence for the existence of time-periodic solutions of the spherically symmetric Einstein-massless scalar field equations with negative cosmological constant in $d+1$ dimensions for any $d\geq 2$. We construct these solutions using two independent methods: nonlinear perturbative expansion  and fully nonlinear numerical evolution. The fact that these two methods  produce the same solutions  make us feel confident in our results. In the case of even $d$ the perturbative construction of time-periodic solution can be algorithmized with  the highest order of the expansion limited only by the memory resources of a computer on which our \textit{Mathematica} script is running. The case of odd $d$  is more involved due to incompatibility between the boundary behavior of the eigenmodes of the linearized problem and the solutions of the full system and its perturbative treatment will not be discussed here (time-periodic solution can be still constructed with slightly different method \cite{Santos}). On the other hand our numerical method for constructing time-periodic solutions works for any $d\geq 2$.
\\
After constructing the time-periodic solutions we check our results
through direct numerical evolution of their initial data. The fact that this numerical evolution reproduces those time-periodic solutions serves as  another cross-check of our results but, even more importantly, it is a strong evidence that these time-periodic solutions  are (nonlinearly) stable. Thus, an interesting picture of dynamics in asymptotically AdS space emerges: although AdS space is generically unstable against black hole formation, it possess also islands of stability. This picture was previously advocated in \cite{dhms}.
\\
It is well known that in asymptotically flat spacetimes there are no non-trivial time-periodic solutions \cite{bst1, bst2}. The mechanism that allows for non-trivial time-periodic solution in our case is the lack of dissipation of energy.
\\
The existence of stable time-periodic solutions of the Einstein equations with negative cosmological constant is interesting in its own right, but it would be also fascinating to learn what is the counterpart of their stability islands on the gauge theory side, using AdS/CFT correspondence.
\\
The finite difference code used in \cite{br} was well suited to perform stable, long-time evolution for the initial data that ultimately collapsed to a black hole, but in the case of solutions that stay smooth forever (like time-periodic solutions) spectral methods are more efficient and moreover they also allow for a direct comparison with perturbative construction of time-periodic solutions. Thus, we construct a pseudo-spectral code that is well suited to deal with the problem at hand.
\\
Finally let us mention that the time-periodic solutions we construct form a particular class of time-periodic solutions with one dominant mode. Whether other time-periodic solutions, that do not bifurcate from one-mode solutions of the linearized problem, exist or not is an open question.
\\
In this note we deal with the simplest model case of the self-gravitating massless scalar field, but the methods presented here can be also applied (with minor modifications) to the pure vacuum case, at least for the cohomogeneity-two Bianchi IX ansatz in \textit{even} number of spatial dimensions, discussed in \cite{bcs}.

\vskip 0.1cm \noindent \emph{Model.} To make this note self-contained we rewrite the equations for the selgravitating massless scalar field with negative cosmological constant from \cite{br,jrb}. We parametrize the $(d+1)$--dimensional asymptotically AdS metric by the ansatz
\begin{equation}
\label{adsd+1:ansatz}
ds^2\! =\! \frac {\ell^2}{\cos^2{\!x}}\left( -A e^{-2 \delta} dt^2 + A^{-1} dx^2 + \sin^2{\!x} \,  d\Omega^2_{d-1}\right),
\end{equation}
where $\ell^2=-d(d-1)/(2\Lambda)$, $d\Omega^2_{d-1}$ is the round metric on $S^{d-1}$, $-\infty<t<\infty$, $0\leq x<\pi/2$, and $A$, $\delta$ are functions of $(t,x)$. For this ansatz the evolution of a self-gravitating massless scalar field $\phi(t,x)$ is governed by the following system (using units where $8\pi G=d-1$)
\begin{equation}
\label{ms_in_ads_d+1:eq_wave}
\dot\Phi = \left( A e^{-\delta} \Pi \right)', \quad \dot \Pi = \frac{1}{\tan^{d-1}{\!x}}\left(\tan^{d-1}{\!x} \,A e^{-\delta} \Phi \right)',
\end{equation}
\begin{align}
\label{ms_in_ads_d+1:eq_00}
A' \!&= \!\frac{d-2+2\sin^2{\!x}} {\sin{x}\cos{x}} \, (1-A) - \sin{x}\cos{x} \, A \left( \Phi^2 + \Pi^2 \right),
\\
\label{ms_in_ads_d+1:eqs_10_11}
\delta' \!&=\! -  \sin{x}\cos{x} \left( \Phi^2 + \Pi^2 \right),
\end{align}
where ${}^{\cdot}=\partial_t$, ${}'=\partial_x$, and
\begin{equation}
\label{Phi_Pi_definitions}
\Phi= \phi', \qquad \Pi= A^{-1} e^{\delta} \dot \phi \,.
\end{equation}
The key point is to note that in this particular gauge both constraint equations (\ref{ms_in_ads_d+1:eq_00}, \ref{ms_in_ads_d+1:eqs_10_11}) can be put in the form that allows for efficient integration. This idea was previously exploited in \cite{m} for the efficient parallelization in solving the constraints in finite difference code, but here we will use this form to construct a spectral code and to find the time-periodic solutions of the system both perurbatively and numerically. For a given matter content, prescribed with the functions $\Phi$ and $\Pi$ treated as independent dynamical variables, the metric function $\delta$ can be expressed as the integral
\begin{equation}
\delta = - \int _{x_0}^x \sin y\, \cos y\, \left( \Phi^2(t,y) + \Pi^2(t,y) \right)\, dy \, ,
\label{delta_constraint_solution}
\end{equation}
where the value of $x_0$ reflects the residual gauge choice in (\ref{adsd+1:ansatz}). The choice $x_0=0$ makes the time coordinate $t$ to be the proper time of the central observer. Now, using (\ref{ms_in_ads_d+1:eqs_10_11}) and the identity
\begin{equation}
\frac {d-2+2\sin^2x} {\sin x \cos x} = \frac {\cos^d x} {\sin^{d-2} x} \left( \frac {\sin^{d-2} x} {\cos^d x} \right)'\,,
\label{integrating_factor}
\end{equation}
we can rewrite (\ref{ms_in_ads_d+1:eq_00}) as
\begin{equation}
\left( \frac {\sin^{d-2} x} {\cos^d x} A \right)' - \left( \frac {\sin^{d-2} x} {\cos^d x} A \right) \delta' = \left( \frac {\sin^{d-2} x} {\cos^d x} \right)' \,.
\label{A_constraint_reformulated}
\end{equation}
Multiplying this equation by $e^{-\delta}$ and integrating by parts we get
\begin{align}
\label{A_constraint_solution}
& 1 - A = e^{\delta} \frac {(\cos x)^d} {(\sin x)^{d-2}} \\
& \times \int_0^x e^{-\delta(t,y)} \left( \Phi^2(t,y) + \Pi^2(t,y) \right) (\tan y)^{d-1}\, dy \, . \nonumber
\end{align}
It is also convenient to rewrite the wave equation (\ref{ms_in_ads_d+1:eq_wave}) eliminating $A'$ and $\delta'$ in the equation for $\dot \Pi$, using the constraint equations (\ref{ms_in_ads_d+1:eq_00}, \ref{ms_in_ads_d+1:eqs_10_11}). In this way we get
\begin{equation}
\label{dot_Pi}
\dot \Pi = e^{-\delta} \left( \Phi' \, A + \Phi \, \frac {d-1-(1-A) \cos 2x} {\sin x \, \cos x} \right).
\end{equation}
The set of equations (\ref{dot_Pi}, \ref{Phi_Pi_definitions}, \ref{delta_constraint_solution}, \ref{A_constraint_solution}) forms a closed system in the form well suited for the numerical integration.

\vskip 0.1cm \noindent \emph{Linear perturbations.}
The spectrum of the linear self-adjoint operator, which governs linearized perturbations of AdS${}_{d+1}$, $L = -(\tan x)^{1-d} \partial_x \left((\tan x)^{d-1} \partial_x \right)$, is given by $\omega_j^2=(d+2j)^2$, $j=0,1,\ldots\,$. The eigenfunctions read
\begin{equation}
\label{eigen_modes}
e_j(x) = 2 \frac {\sqrt{j!\, (j+d-1)!}} {\Gamma\left(j + \frac {d}{2}\right)} (\cos x)^d P_j^{d/2-1,d/2} (\cos 2x)\,,
\end{equation}
where $P_j^{\alpha,\beta} (x)$ are the Jacobi polynomials. These eigenfunctions form an orthonormal base in the Hilbert space of functions $L^2\left([0,\pi/2], \, (\tan x)^{d-1} \, dx\right)$. Below we denote the inner product on this Hilbert space by $\inner{f}{g} := \int _0 ^{\pi/2} f(x) g(x) (\tan x)^{d-1}\, dx$. It will be useful in the following to note that $\inner{e_i'(x)}{e_j'(x)} = \omega_j^2 \delta_{ij}$.

\vskip 0.1cm \noindent \emph{Perturbative construction of time-periodic solution.}
We seek a time-periodic solution of the system (\ref{ms_in_ads_d+1:eq_wave}-\ref{ms_in_ads_d+1:eqs_10_11}) in the form
\begin{align}
\phi \!&= \ep \, \cos (\tau) e_{\gamma}(x)+ \sum_{\mbox{{\small odd }} \lambda \geq 3} \ep^{\lambda} \, \phi_{\lambda}(\tau, x),
\label{pexp_phi}
\\
\delta \!&=\sum_{\mbox{{\small even }} \lambda \geq 2} \ep^{\lambda} \, \delta_{\lambda}(\tau, x), \qquad 1-A = \sum_{\mbox{{\small even }} \lambda \geq 2} \ep^{\lambda} \, A_{\lambda}(\tau, x),
\label{pexp_delta_A}
\end{align}
where $e_{\gamma}(x)$ is a \textit{dominant} mode in the solution in the limit $\ep \ra 0$, $\tau =\Omega_{\gamma} t$ is the rescaled time variable with
\begin{equation}
\label{Omega}
\Omega_{\gamma} = \omega_{\gamma} + \sum_{\mbox{{\small even }} \lambda \geq 2} \ep^{\lambda} \, \omega_{\gamma,\lambda}
\end{equation}
and
\begin{align}
\label{pexp_phi_2}
\phi_{\lambda} \!&= \sum_{j} f_{\lambda,j}(\tau) e_j(x),
\\
\label{pexp_delta_A_2}
\delta_{\lambda} \!&= d_{\lambda,-1}(\tau) + \sum_j d_{\lambda,j}(\tau) e_j(x), \quad A_{\lambda} = \sum_j a_{\lambda,j}(\tau) e_j(x) \, ,
\end{align}
with $f_{\lambda,j}(\tau)$, $a_{\lambda,j}(\tau)$, $d_{\lambda,j}(\tau)$ being periodic in $\tau$. It is important to note that for even $d$ the sums in (\ref{pexp_phi_2}, \ref{pexp_delta_A_2}) are finite at each order $\lambda$ of the perturbative expansions (\ref{pexp_phi}, \ref{pexp_delta_A}). This allows for a straightforward algorithmization of building up the successive terms in (\ref{pexp_phi}, \ref{pexp_delta_A}). That is not the case for odd $d$. This is due to the boundary expansion of the solutions of the system (\ref{ms_in_ads_d+1:eq_wave}-\ref{ms_in_ads_d+1:eqs_10_11}). Smoothness at spatial infinity and finiteness of the total mass $M$ imply that near $x=\pi/2$ we must have (using $\rho=\pi/2-x$)
\begin{align}
\label{pi2}
\phi(t,x) \!&= \mathcal{O}(\rho^{d}),
\nonumber \\
A(t,x)\!&=1+\mathcal{O}(\rho^{d}), \quad \delta(t,x)=\mathcal{O}(\rho^{2 d})\,.
\end{align}
It can be checked that for odd $d$, higher even terms in the expansion of $\phi$ at the boundary $x=\pi/2$ do not vanish \footnote{They vanish only in the limit $M \ra 0$.} and $\phi_{\lambda}$ does not have finite expansion into eigenmodes $e_j(x)$, being (in odd $d$) odd, with respect to $x=\pi/2$. For later convenience it is useful to introduce the following notation: let $\coef{\lambda} f$ denote the coefficient at $\ep^{\lambda}$ in the (formal) power series expansion of $f=\sum_{\lambda} f_{\lambda} \ep^{\lambda}$. Now, inserting the first of the series in (\ref{pexp_delta_A}, \ref{pexp_delta_A_2}) into (\ref{ms_in_ads_d+1:eqs_10_11}) and projecting onto $e_k'(x)$ we get
\begin{equation}
d_{\lambda, k} = - \frac{1}{2 \omega_k^2} \inner{e_k'}{\coef{\lambda} \sin(2x) \left( \Phi^2 + \Pi^2 \right)}
\end{equation}
and $d_{\lambda, -1}(\tau)$ is fixed by the gauge fixing condition $\delta_{\lambda}(\tau,0)=0$.
Similarly, inserting the last of the series in (\ref{pexp_delta_A}, \ref{pexp_delta_A_2}) into (\ref{A_constraint_reformulated}) using (\ref{ms_in_ads_d+1:eqs_10_11}, \ref{integrating_factor}) and then projecting onto $e_k(x)$ we get a linear system of equations for the coefficients $a_{\lambda,j}(\tau)$:
\begin{align}
& \sum_j \left[ (d-1) \delta_{kj} + \inner{e_k}{\frac {1}{2} \sin2x \, e_j' - \cos 2x\, e_j}\right] a_{\lambda,j} =
\nonumber\\
& \frac{1}{4} \inner{e_k}{\coef{\lambda}(\sin 2x)^2 A \left(\Phi^2 + \Pi^2\right)} \,.
\end{align}
It is useful to note that the principal matrix of this system is 3-diagonal. This system supplied with $\left. \coef{\lambda}(1-A) \right|_{x=0} = 0 = \sum_j a_{\lambda,j} e_j(0)$ condition allows for the unique solution for the coefficients $a_{\lambda,j}(\tau)$. Then, for odd $\lambda \geq 3$, one gets that $\phi_{\lambda}$ fulfills an inhomogeneous wave equation on the pure AdS background: $\left(\omega_{\gamma}^2 \partial_{\tau\tau} + L\right) \phi_{\lambda} = S_{\lambda}$. Projecting this equation onto $e_k$ one finds that the coefficients $f_{\lambda,k}$ in (\ref{pexp_phi_2}) behave as forced harmonic oscillators: $\left(\omega_{\gamma}^2 \partial_{\tau\tau} + \omega_{k}^2\right) f_{\lambda,k} = \inner{e_k}{S_{\lambda}}$. Solving these, we get two integration constants for each of the equations. The two constants in $f_{\lambda,\gamma}$ (the coefficient for the \textit{dominant} mode $e_{\gamma}$ in (\ref{pexp_phi_2})), are fixed with the condition $\left. \left( f_{\lambda, \gamma}, \, \partial_{\tau} f_{\lambda, \gamma} \right) \right|_{\tau=0} = (0,\, 0)$, that is we choose $\left. \left( \inner{e_{\gamma}}{\phi}, \, \inner{e_{\gamma}}{\partial_{\tau} \phi} \right) \right|_{\tau=0} = (\ep, \, 0)$. Now, if $\inner{e_k}{S_{\lambda}}$ contains the resonant terms $\cos (\omega_{k}/\omega_{\gamma}) \tau$ or $\sin (\omega_{k} / \omega_{\gamma}) \tau$, this gives rise to secular terms $\tau \sin (\tau \, \omega_{k} / \omega_{\gamma})$ and $\tau \cos (\tau \, \omega_{k} / \omega_{\gamma})$ in $f_{\lambda,k}$ respectively. Such terms would spoil the periodicity of the solution, thus they have to be removed. This fixes the correction to the frequency $\omega_{\gamma, \lambda-1}$, and the integration constants. Namely, it turns out that in order to not to produce spurious resonant terms in higher orders, all but odd (in $\tau$) frequencies in the solutions for $f_{\lambda,k}$ have to be removed and moreover, all $f_{\lambda,k}$ tune in phase to the dominant mode: $\left. \partial_{\tau} f_{\lambda,k} \right|_{\tau=0}=0$ (for the choice $\left. \partial_{\tau} f_{\lambda, \gamma} \right|_{\tau=0}=0$). In summary, at any odd $\lambda \geq 3$ we fix $\omega_{\gamma,\lambda-1}$ and, as $\inner{e_k}{S_{\lambda}}$ does not vanish only for a finite number of modes (namely $\inner {e_k} {S_{\lambda}} \equiv 0$ for $k>\gamma + (d+1+2\gamma)(\lambda-1)/2$), we are left with $(\lambda-1)/2 + \lfloor (\lambda-1)/(2 (d+2\gamma))\rfloor$ undetermined integration constants. They will be fixed, together with $\omega_{\gamma,\lambda+1}$, to remove $(\lambda+1)/2 + \lfloor (\lambda-1)/(2 (d+2\gamma))\rfloor$ secular terms in $\phi_{\lambda+2}$.

In fact all the projections onto $e_k$ (or $e'_k$) appearing at any order of the perturbative procedure described above, can be reduced to just a few inner products: $\inner {e_k} {e_i \, e_j}$, $\inner {e_k} {\cos 2x \, e_i}$, $\inner {e_k} {\sin 2x \, e_i'}$, $\inner {e_k} {\csc x \, \sec x \, \cos 2x \, e_i \, e_j'}$, $\inner {e_k} {\csc x \, \sec x \, e_i \, e_j'}$. Thus, the whole procedure of building up such perturbative solution is straightforward to implement.

\vskip 0.1cm \noindent \emph{Pseudo-spectral code for the time evolution.}
We expand $\phi$ and $\Pi$ into $K$ eigenmodes of linearized problem (\ref{eigen_modes})
\begin{equation}
\phi = \sum _{0 \leq j < K} f_j(t) e_j(x)\,, \qquad \Pi = \sum _{0 \leq j < K} p_j(t) e_j(x)
\label{spectral_decomposition_dynamics}
\end{equation}
Prescribing the coefficients $f_j(t)$, $p_j(t)$ at time $t$, we want to solve for them at a later time $t+dt$. To achieve that, we calculate their time derivatives $\dot f_j(t)$ and $\dot p_j(t)$ with a spectral method and then integrate them in time. We know that the metric function $\delta$ and the integrand in (\ref{A_constraint_solution}) can be efficiently \footnote{That is the expansion coefficient decay exponentially for smooth dynamical fields $\phi$ and $\Pi$.} expanded as follows
\begin{align}
\delta \!&=\! \sum _{0 \leq j < K} d_j(t) \cos(2 j x) \,,
\label{spectral_decomposition_delta_constraint}
\\
e^{-\delta} \left(\Phi^2 + \Pi^2\right) \!&=\! \sum _{0 \leq j < K} u_j(t) \cos(2 j x) \,.
\label{spectral_decomposition_A_constraint}
\end{align}
Substituting (\ref{spectral_decomposition_delta_constraint}) into (\ref{ms_in_ads_d+1:eqs_10_11}) we get
\begin{equation}
\sum _{1 \leq j < K} (-2 j) \sin(2 j x) d_j(t) = -\sin x \, \cos x \left(\Phi^2 + \Pi^2\right)\,.
\label{spectral_decomposition_delta_prime}
\end{equation}
When the sides of (\ref{spectral_decomposition_delta_prime}) are evaluated at $K$ collocation points, $x_k = (\pi/2) k / (K+1)$, $k=1,\ldots,K$, we get the linear system of equations to be solved for the coefficients $d_j(t)$. As $d_0(t)$ is absent in this system, reflecting the gauge freedom, we add the gauge fixing condition
\begin{equation}
  \delta(t,0) = \sum _{0 \leq j < K} d_j(t) = 0\,.
\end{equation}
Then, evaluating the sides of (\ref{spectral_decomposition_A_constraint}) at the collocation points we get the linear system of equations to be solved for the coefficients $u_j(t)$. This allows us to solve for the metric function $A$. Using (\ref{A_constraint_solution}) we get
\begin{equation}
A = 1 - e^{\delta} \frac {(\cos x)^d} {(\sin x)^{d-2}} \sum _{0 \leq j < K} w^{(d)}_j(x) u_j(t)\,,
\label{metric_function_A_solution}
\end{equation}
where the weight functions $w^{(d)}_j(x)$ read:
\begin{equation}
w^{(d)}_j(x) = \int_0^{x} \cos(2 j y) (\tan y)^{d-1} \, dy\,.
\label{weights}
\end{equation}
Now, substituting the expansions (\ref{spectral_decomposition_dynamics}) into wave equation (\ref{dot_Pi}, \ref{Phi_Pi_definitions}) and evaluating both sides at the collocation points we get the linear system of equations to be solved for the time derrivatives $\dot f_j(t)$ and $\dot p_j(t)$.

\vskip 0.1cm \noindent \emph{Numerical construction of time-periodic solutions.}
Seeking for time-periodic solutions numerically it is convenient to use rescaled time coordinate $\tau=\Omega t$ where, as in the perturbative construction, $\Omega$ denotes the frequency of the solution we are looking for. Assuming that time-periodic solution does exist, we expand $\phi(t,x)$ and $\Pi(t,x)$ into eigenmodes of the linearized problem in space and Fourier coefficients in time. Choosing a grid with $K$ collocation points $x_k = (\pi/2) k / (K+1)$, $k=1,\ldots,K$ in space (as for the spectral code) and $N$ collocation points in time $\tau_n = \pi(n-1/2)/(2N+1)$, $n=1,\ldots,N$ we truncate these expansions as follows
\begin{align}
\phi \!&= \sum _{0 \leq i < N} \sum _{0 \leq j < K} f_{i,j} \cos((2i+1) \tau) e_j(x)\,,
\\
\Pi \!&= \sum _{0 \leq i < N} \sum _{0 \leq j < K} p_{i,j} \sin((2i+1) \tau) e_j(x) \,.
\label{spectral_decomposition_numerics}
\end{align}
Next, at each instant of time $\tau_n$ we calculate the coefficients
\begin{align}
f_j(t_n) \!&= \sum _{0 \leq i < N} f_{i,j} \cos((2i+1) \tau_n) \,,
\\
p_j(t_n) \!&= \sum _{0 \leq i < N} p_{i,j} \sin((2i+1) \tau_n) \,
\label{spectral_decomposition_numerics_tau_n}
\end{align}
and put them as an input to our spectral procedure, getting as the output their time derivatives $\dot f_j(t_n)$ and $\dot p_j(t_n)$. Equating those to the time derrivatives of (\ref{spectral_decomposition_numerics}) (remembering that $\partial_t = \Omega \partial_{\tau}$) at the set of $K \times N$ grid points $(\tau_n, x_k)$, $n=1,\ldots,N$, $k=1,\ldots,K$, together with (\ref{spectral_decomposition_numerics}) inserted into the second of the equations (\ref{Phi_Pi_definitions}) and then evaluated at the set of $K \times N$ grid points, and the equation $\sum _{0 \leq i < N} f_{i,\gamma} = \ep$, setting the amplitude of the dominant mode $\gamma$ in the initial data to $\ep$, we get a nonlinear system of $2 \times K \times N + 1$ equations for $2 \times K \times N + 1$ unknowns: $f_{i,j}$, $p_{i,j}$ and $\Omega$. This system is solved with Newton-Raphson algorithm yielding the time-periodic solution for the system (\ref{ms_in_ads_d+1:eq_wave}-\ref{ms_in_ads_d+1:eqs_10_11}).

\begin{table}[h]
  \centering
  \begin{tabular}{c|l||c|l||c|l}
    $j$ & \multicolumn{1}{c||}{$f_{j}(0)$} & $j$ & \multicolumn{1}{c||}{$f_{j}(0)$} & $j$ & \multicolumn{1}{c}{$f_{j}(0)$} \\ \hline\hline
    0 &	\phantom{$-$}$0.01$                 &     5 &	$6.82972\cdot 10^{-9}$ &     10 &	$1.95754\cdot 10^{-14}$ \\
    1 &	\phantom{$-$}$1.04021\cdot 10^{-5}$ &     6 &	$4.91226\cdot 10^{-11}$ &     11 &	$3.26051\cdot 10^{-16}$ \\
    2 &	\phantom{$-$}$1.68282\cdot 10^{-6}$ &     7 &	$6.71227\cdot 10^{-12}$ &  \multirow{2}{*}{$\vdots$}  & \multirow{2}{*}{$\vdots$}  \\
    3 &	\phantom{$-$}$2.09044\cdot 10^{-7}$ &     8 &	$8.72677\cdot 10^{-13}$ &  &  \\
    4 &	$-1.65282\cdot 10^{-9}$&     9 &	$2.74313\cdot 10^{-14}$ &     63 & $2.31645\cdot 10^{-69}$ \\
  \end{tabular}
  \caption{Coefficients of the expansion (\ref{spectral_decomposition_dynamics}) at $t=0$
    (with~$p_j(0) \equiv 0$), corresponding to the time-periodic solution with dominant
    eigenmode $\gamma=0$, with the amplitude $\ep=1/100$, determined by the numerical procedure
    with $K\times N=64\times 24$ grid points and extended floating-point precision.
    The solution oscillates with frequency $\Omega\approx 4.00667$.}
  \label{tab:initial_data}
\end{table}

\vskip 0.1cm \noindent \emph{Results.} The detailed results obtained with the tools described above will be given in a longer accompanying paper. Here, just to give an example, we present the time-periodic solution bifurcating from the fundamental mode (as a dominant one) in $4+1$ dimensions. First, we construct the time-periodic solution with a numerical procedure described above.
\begin{figure}[h]
  \centering
  \includegraphics[width=0.99\columnwidth]{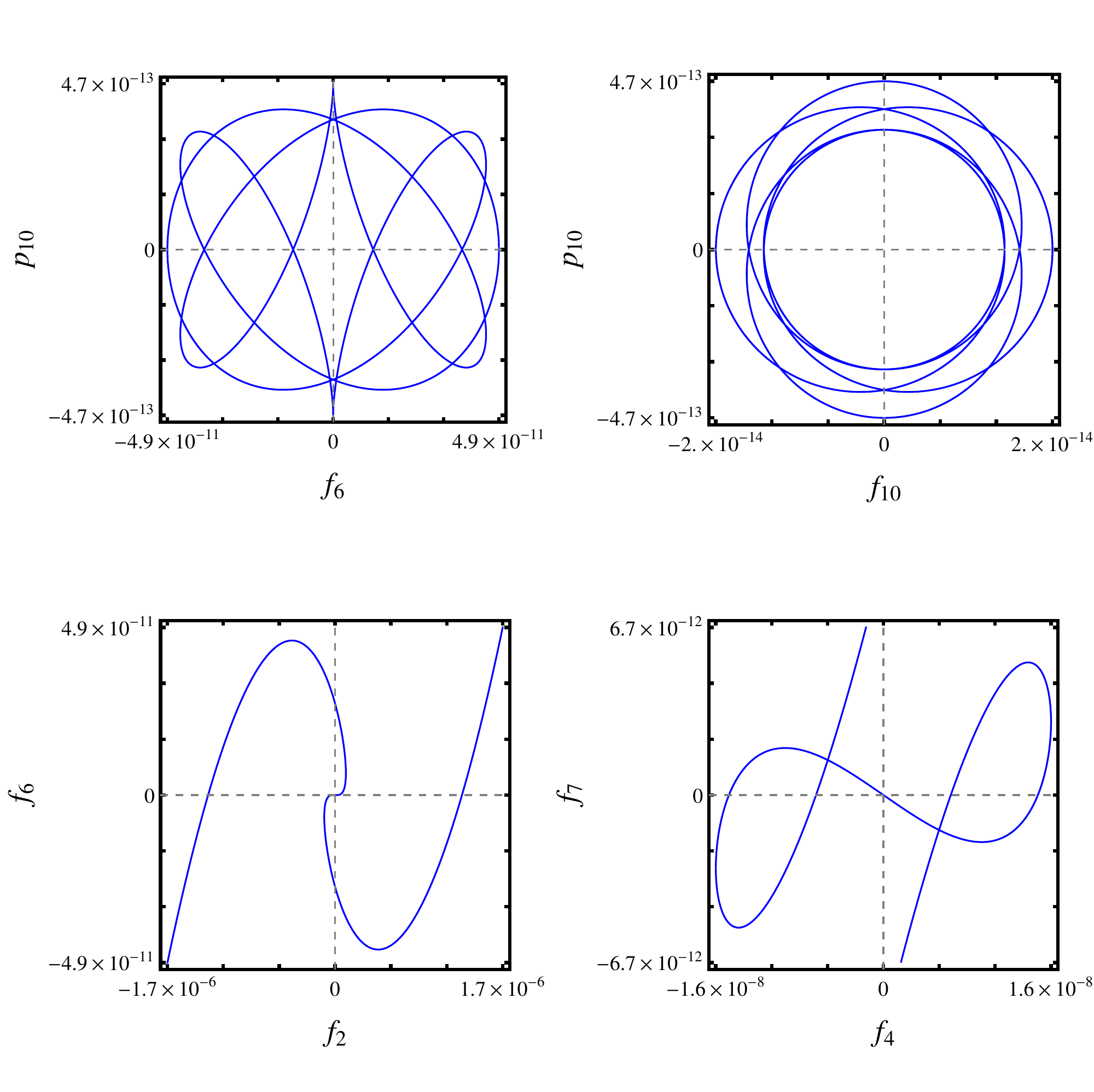}
  \caption{Visualization of the phase space generated by time
    evolution of initial data listed in
    Tab.~\ref{tab:initial_data}. All plots depict the
    same solution evolved over time interval equal to 500
    periods. Closed curves on the slices of phase space give a strong
    argument for the stability of analyzed solution.}
  \label{fig:phase_plot}
\end{figure}
Then, we read off the coefficients of the expansion (\ref{spectral_decomposition_dynamics}) at $t=0$ (see Tab.~\ref{tab:initial_data}) and put them as the initial data into our spectral evolution code. Despite the presence of truncation errors and numerical noise in these initial data, their time evolution is periodic in time, as depicted by closed loops in Fig.~\ref{fig:phase_plot} representing different sections of the phase space, spanned by the set of coefficients $\{f_j(t), \, p_k(t)\}$. This provides strong evidence not only for the existence of the time-periodic solutions but also for their (nonlinear) stability. This argument for the stability is strengthen by the fact that if we perturb the time-periodic solution slightly then its evolution is no longer periodic, but it does not collapse to a black hole and stays close to the periodic orbit. Using the perturbative method, for small values of $\ep$ we get fast convergence to the numerical solution, for larger $\ep$ the convergence can be improved with the Pad\'e resummation (see Tab.~\ref{tab:omega_compare}). The Pad\'e approximation can be also used to estimate the convergence radius for perturbative expansions (\ref{pexp_phi}-\ref{Omega}). For example, if we construct the $(n,n)$ Pad\'e approximants for
\begin{widetext}
\begin{align}
  \label{eq:perturbative_freq}
\begin{split}
  \Omega_{\gamma=0} & = 4+\frac{464}{7}\,\ep^2+\frac{45614896}{11319}\,\ep^4 +\frac{173158711507904383595696}{533797475350414275}\,\ep^6 +\frac{19627018631453126466665156076805265104}{662148921092395909349993941125}\,\ep^8 \\[0.8ex]
  &+\frac{11072083972904297030696081837311640731963665649719830983726203187072}{3798775287765327851156088282697847314054859440405429239375}\,\ep^{10} \\[0.8ex]
  &+\frac{20613105125948158448048760017030229079689748556646392089437804474799052885011038708688}{68546459749514836602259551946772532940114419607536660561179461719235846875}\,\ep^{12} \\[0.8ex]
  &+ \omega_{0,14}\,\ep^{14} + \omega_{0,16}\,\ep^{16} + \mathcal{O}\left(\ep^{18}\right),
\end{split}
\end{align}
\end{widetext}
with $n=2,4,6,8$ \footnote{The terms $\omega_{0,14}$, $\omega_{0,16}$ have been calculated but their numerators are 124, 153 - digits numbers respectively, so we do not show them explicitly.} and then calculate the zeros of the denominators nearest to the origin we get $0.128$, $0.102$, $0.095$ and $0.092$, respectively. These values are consistent with the fact that the Newton-Raphson algorithm used in the numerical construction of time-periodic solutions ceases to converge for $\ep \gtrsim 0.09$. Moreover if we read off the initial data for larger values of $\ep$ the time evolution leads to a black hole formation almost immediately.

\begin{table}[th]
  \centering
  \begin{tabular}{c | l l l}
 $\ep$ & \multicolumn{1}{c}{$\Omega _{\Sigma}$} & \multicolumn{1}{c}{$\Omega _{\text{Pad\'e}}$} & \multicolumn{1}{c}{$\Omega _{\text{num}}$}  \\ \hline\hline
 0.005 & 4.0016596666501 & 4.0016596666501 & 4.0016596666501 \\
 0.015 & 4.0151220741462 & 4.0151220741462 & 4.0151220741462 \\
 0.025 & 4.0430867838460 & 4.0430867838521 & 4.0430867838521 \\
 0.035 & 4.0879197007 & 4.0879197035435 & 4.0879197035448 \\
 0.045 & 4.15407139 & 4.15407167953 & 4.1540716797440 \\
 0.055 & 4.249920 & 4.249932516 & 4.2499325336279 \\
 0.065 & 4.39267 & 4.3929928 &  4.3929938556099 \\
 0.075 & 4.6230 & 4.629225 & 4.6292962269712 \\
 0.085 & 5.05 & 5.18 & 5.2017714694183 \\ \hline
  \end{tabular}
  \caption{Frequency of solutions with $\gamma=0$ determined by:
    direct summation of the perturbative expansion $\Omega_{\Sigma}$
    \eqref{eq:perturbative_freq}, diagonal (8,8) Pad\'e resumation
    $\Omega_{\text{Pad\'e}}$ and the numerical code $\Omega _{\text{num}}$.
    The relative absolute difference of $\Omega _{\text{num}}$ and $\Omega_{\text{Pad\'e}}$
    ranges from $10^{-28}$ for $\ep=0.005$ to $4\cdot 10^{-3}$ for $\ep=0.085$.}
  \label{tab:omega_compare}
\end{table}

\vskip 0.1cm \noindent \emph{Acknowledgments:} We are indebted to
Piotr Bizo\'n for suggestions and discussions. This work was supported
by the NCN grant DEC-2012/06/A/ST2/00397.


\begin{thebibliography}{10}

\bibitem{br}
P. Bizo\'n and A. Rostworowski,
%
Phys. Rev. Lett. \textbf{107}, 031102 (2011), arXiv:1104.3702

\bibitem{jrb}
J. Ja\l{}mu\.zna, A. Rostworowski and P. Bizo\'n,
Phys. Rev. D \textbf{84}, 085021 (2011), arXiv:1108.4539

\bibitem{bll}
A. Buchel, L. Lehner and S.L. Liebling,
Phys. Rev. D \textbf{86}, 123011 (2012), arXiv:1210.0890

\bibitem{dhs}
O.J.C. Dias, G.T. Horowitz and J.E. Santos,
Class. Quant. Grav. \textbf{29}, 194002 (2012), arXiv:1109.1825

\bibitem{dhms}
O.J.C. Dias, G.T. Horowitz, D. Marolf and J.E. Santos,
Class. Quant. Grav. \textbf{29}, 235019 (2012), arXiv:1208.5772

\bibitem{Santos}
J.E. Santos, \textit{private communication:} perturbative construction of the time-periodic solution for a self-gravitating massless scalar field in $3+1$ at spherical symmetry up to $5$-th order of perturbative expansion.

\bibitem{bst1}
J. Bi\v{c}\'ak, M. Scholtz and P. Tod,
Class. Quant. Grav. \textbf{27}, 055007 (2010), arXiv:1003.3402

\bibitem{bst2}
J. Bi\v{c}\'ak, M. Scholtz and P. Tod,
Class. Quant. Grav. \textbf{27}, 175011 (2010), arXiv:1008.0248

\bibitem{bcs}
P. Bizo\'n, T. Chmaj, and B.G. Schmidt, Phys. Rev. Lett.~\textbf{95}, 071102 (2005), arXiv:gr-qc/0506074

\bibitem{m}
M. Maliborski,
Phys. Rev. Lett. \textbf{109}, 221101 (2012), arXiv:1208.2934

\end{thebibliography}
\end{document}